\documentclass{article}%
\usepackage{makeidx}
\usepackage{amsmath}
\usepackage{amsfonts}
\usepackage{amssymb}
\usepackage{graphicx}%
\providecommand{\U}[1]{\protect\rule{.1in}{.1in}}

\begin{document}

\author{Giuseppe Castagnoli
\and Pieve Ligure, Italy, giuseppe.castagnoli@gmail.com}
\title{Quantum computation and the physical computation level of biological
information processing }
\maketitle

\begin{abstract}
On the basis of introspective analysis, we establish a crucial requirement for
the physical computation basis of consciousness: it should allow processing a
significant amount of information together at the same time. Classical
computation does not satisfy the requirement. At the fundamental physical
level, it is a network of two body interactions, each the input-output
transformation of a universal Boolean gate. Thus, it cannot process together
at the same time more than the three bit input of this gate -- many such gates
in parallel do not count since the information is not processed together.
Quantum computation satisfies the requirement. At the light of our recent
explanation of the speed up, quantum measurement of the solution of the
problem is analogous to a many body interaction between the parts of a perfect
classical machine, whose mechanical constraints represent the problem to be
solved. The many body interaction satisfies all the constraints together at
the same time, producing the solution in one shot. This shades light on the
physical computation level of the theories that place consciousness in quantum
measurement and explains how informations coming from disparate sensorial
channels come together in the unity of subjective experience. The fact that
the fundamental mechanism of consciousness is the same of the quantum speed
up, gives quantum consciousness a potentially enormous evolutionary advantage.

\end{abstract}

\section{Introduction}

On the basis of the introspective analysis of visual perception, we establish
a crucial requirement for the physical computation basis of consciousness. In
this moment I see the meeting room, the audience, the chairs, a lot of things
"together at the same time". This is an intuitive statement we cannot easily
do without. I do not see the audience and the chairs at different times.
Consciousness concerns the present time. I certainly see the audience and the
chairs together and at the same time. We translate this statement into the
language of information theory. Seeing implies recognizing, thus processing.
Therefore the physical computation basis of consciousness should allow
processing a significant amount of information (at least that of a digital
picture) together at the same time. We translate "together" into impossibility
of breaking down the information processing into independent processings and
assume that "at the same time" is to be be understood in a non-relativistic framework.

We compare classical and quantum computation with the requirement.

At the fundamental physical level, classical computation is represented by a
network of two body interactions, each the input-output transformation of a
universal Boolean gate. The maximum amount of information processed together
at the same time, occurring in the instant of the collision between two
bodies, is the three bit input of the above gate. Many such gates in parallel
do not count since the information is not processed together.

Quantum computation is examined at the light of our recent explanation of the
"quantum speed up" (quantum algorithms requiring less computations than
classical algorithms). Because of retrocausation, 50\% of the information
about the solution of the problem, acquired by measuring the content of the
computer register at the end of the algorithm, goes back in time to before
running the algorithm. The quantum algorithm uses this information to compute
the solution with a lower number of operations. It is a superposition of
causal/local computation histories, each corresponding to a possible way of
getting in advance 50\% of the information about the solution.

This retrocausation mechanism has an idealized classical analog, useful to
compare quantum computation with the requirement. The quantum measurement that
produces the solution is analogous to a many body interaction between the
parts of a perfect classical machine. The classical representation of quantum
retrocausation and nonlocality requires perfect machine rigidity, accuracy,
and reversibility. The mechanical constraints of this machine represent the
logical constraints of the problem to be solved. The many body interaction
senses and satisfies all the constraints together at the same time, producing
the solution in one shot. In contrast, classical computation, processing at
most three bits at the same time, cannot take into account all the problem
constraints at the same time; this leads to trial and error and to the
relative zero of the quantum speed up.

Summing up, quantum computation satisfies the requirement of the physical
computation basis of consciousness, which turns out to be the prerequisite of
the quantum speed up. This shades light on the physical computation level of
the theories that place consciousness in quantum measurement and explains how
informations coming from disparate sensorial channels come together in the
unity of subjective experience. The fact that the fundamental mechanism of
consciousness is the same of the quantum speed up, gives quantum consciousness
a potentially enormous evolutionary advantage.

In the following, after reviewing the quantum database search algorithm, we
provide its many body representation. Then we show that the explanation of the
speed up interplays with a variety of scientific and philosophical issues
concerning consciousness and, more in general, biological information processing.

\section{Reviewing Grover's algorithm}

We review Grover's quantum data base search algorithm in the simple instance
of data base size $N=4$. For the sake of interdisciplinarity, we explain
Grover's algorithm from scratch, without requiring any previous knowledge of
quantum computer science. Data base search is seen as a game between two
players. We have a chest of 4 drawers numbered 00, 01, 10, 11, a ball, and the
two players. The oracle hides the ball in drawer number $\mathbf{k}%
\equiv\mathbf{~}k_{0},k_{1}$ and gives to the second player the chest of
drawers, represented by a black box that, given in input a drawer number
$\mathbf{x}\equiv x_{0},x_{1}$, computes the Kronecker function $f_{\mathbf{k}%
}\left(  \mathbf{x}\right)  =\delta\left(  \mathbf{k},\mathbf{x}\right)  $ (1
if $\mathbf{k}=\mathbf{x}$, 0 otherwise). The second player -- the algorithm
-- should find the number of the drawer with the ball, and this is done by
computing $\delta\left(  \mathbf{k},\mathbf{x}\right)  $ for different values
of $\mathbf{x}$ -- by opening different drawers. A classical algorithm
requires 2.25 computations of $\delta\left(  \mathbf{k},\mathbf{x}\right)
$\ on average, 3 computations if one wants to be a priori certain of finding
the solution. The quantum algorithm yields the solution with certainty with
just one computation.

In our representation of the quantum algorithm, the computer has three
registers. A two-qubit register $K$ contains the oracle's choice of the value
of $\mathbf{k}$. The state $\left\vert 00\right\rangle _{K}$, or $\left\vert
01\right\rangle _{K}$, etc. of this register means oracle's choice
$\mathbf{k}=00$,\ or $\mathbf{k}=01$, etc.; of course the state of any
register can also be a superposition of sharp quantum states. Register $K$\ is
only a useful conceptual reference, it provides a panoramic view of the
behavior of the quantum algorithm for all the possible oracle's choices. Then
there are the two-qubit register $X$ containing the argument $\mathbf{x}$ to
query the black box with and the one-qubit register $V$ meant to contain the
result of the computation, modulo 2 added to its initial content for logical
reversibility. The three registers undergo a unitary evolution, where in
particular $\delta\left(  \mathbf{k},\mathbf{x}\right)  $ is computed once.
Measuring $\left[  K\right]  $, the content of register $K$, yields the
oracle's choice $\mathbf{k}$; this measurement can be performed,
indifferently, at the beginning or at the end of the algorithm -- which is in
fact the identity in the Hilbert space of $K$. Measuring $\left[  X\right]  $
at the end of the algorithm yields the solution of the problem $\mathbf{x}%
=\mathbf{k}$.

The initial state of the three registers is:%

\begin{equation}
\frac{1}{4\sqrt{2}}\left(  \left\vert 00\right\rangle _{K}+\left\vert
01\right\rangle _{K}+\left\vert 10\right\rangle _{K}+\left\vert
11\right\rangle _{K}\right)  \left(  \left\vert 00\right\rangle _{X}%
+\left\vert 01\right\rangle _{X}+\left\vert 10\right\rangle _{X}+\left\vert
11\right\rangle _{X}\right)  \left(  \left\vert 0\right\rangle _{V}-\left\vert
1\right\rangle _{V}\right)  . \label{input}%
\end{equation}
Preparing $K$ in a uniform superposition of the four possible oracle's choices
provides a panoramic view of the behavior of the quantum algorithm. We can
switch to a single choice by measuring $\left[  K\right]  $\ in (\ref{input}),
also after having prepared $K$ in a desired sharp quantum state (for
uniformity of language, we see a classical preparation of $K$ as a measurement outcome).

State (\ref{input}) is the input of the computation of $\delta\left(
\mathbf{k},\mathbf{x}\right)  $, which is performed in quantum parallelism on
each term of the superposition. E. g. the input term $-\left\vert
01\right\rangle _{K}\left\vert 01\right\rangle _{X}\left\vert 1\right\rangle
_{V}$ means that the input of the black box is $\mathbf{k}=01,$ $\mathbf{x}%
=01$ and that the initial content of register $V$\ is 1. The computation
yields $\delta\left(  01,01\right)  =1$, which modulo 2 added to the initial
content of $V$\ yields the output term $-\left\vert 01\right\rangle
_{K}\left\vert 01\right\rangle _{X}\left\vert 0\right\rangle _{V}$ ($K$\ and
$X$ keep the memory of the input for logical reversibility). Similarly, the
input term $\left\vert 01\right\rangle _{K}\left\vert 01\right\rangle
_{X}\left\vert 0\right\rangle _{V}$ goes into the output term $\left\vert
01\right\rangle _{K}\left\vert 01\right\rangle _{X}\left\vert 1\right\rangle
_{V}$. Summing up, $\left\vert 01\right\rangle _{K}\left\vert 01\right\rangle
_{X}\left(  \left\vert 0\right\rangle _{V}-\left\vert 1\right\rangle
_{V}\right)  $ goes into $-\left\vert 01\right\rangle _{K}\left\vert
01\right\rangle _{X}\left(  \left\vert 0\right\rangle _{V}-\left\vert
1\right\rangle _{V}\right)  $. The computation of $\delta\left(
\mathbf{k},\mathbf{x}\right)  $\ inverts the phase of those $\left\vert
\mathbf{k}\right\rangle _{K}\left\vert \mathbf{x}\right\rangle _{X}\left(
\left\vert 0\right\rangle _{V}-\left\vert 1\right\rangle _{V}\right)  $ where
$\mathbf{k}=\mathbf{x}$ and is the identity otherwise. In the overall, it
changes (\ref{input}) into:%

\begin{equation}
\frac{1}{4\sqrt{2}}\left[
\begin{array}
[c]{c}%
\left\vert 00\right\rangle _{K}\left(  -\left\vert 00\right\rangle
_{X}+\left\vert 01\right\rangle _{X}+\left\vert 10\right\rangle _{X}%
+\left\vert 11\right\rangle _{X}\right)  +\\
\left\vert 01\right\rangle _{K}\left(  \left\vert 00\right\rangle
_{X}-\left\vert 01\right\rangle _{X}+\left\vert 10\right\rangle _{X}%
+\left\vert 11\right\rangle _{X}\right)  +\\
\left\vert 10\right\rangle _{K}\left(  \left\vert 00\right\rangle
_{X}+\left\vert 01\right\rangle _{X}-\left\vert 10\right\rangle _{X}%
+\left\vert 11\right\rangle _{X}\right)  +\\
\left\vert 11\right\rangle _{K}\left(  \left\vert 00\right\rangle
_{X}+\left\vert 01\right\rangle _{X}+\left\vert 10\right\rangle _{X}%
-\left\vert 11\right\rangle _{X}\right)
\end{array}
\right]  \left(  \left\vert 0\right\rangle _{V}-\left\vert 1\right\rangle
_{V}\right)  , \label{secondstage}%
\end{equation}
a maximally entangled state where four orthogonal states of $K$\ , each
corresponding to a single value of $\mathbf{k}$, are correlated with four
orthogonal states of $X$. This means that the information about the value of
$\mathbf{k}$\ has propagated to $X$.

A suitable rotation of the\ measurement basis of $X$ transforms entanglement
between $K$\ and $X$\ into correlation between the outcomes of measuring their
contents, transforming (\ref{secondstage}) into:%

\begin{equation}
\frac{1}{2\sqrt{2}}\left(  \left\vert 00\right\rangle _{K}\left\vert
00\right\rangle _{X}+\left\vert 01\right\rangle _{K}\left\vert 01\right\rangle
_{X}+\left\vert 10\right\rangle _{K}\left\vert 10\right\rangle _{X}+\left\vert
11\right\rangle _{K}\left\vert 11\right\rangle _{X}\right)  \left(  \left\vert
0\right\rangle _{V}-\left\vert 1\right\rangle _{V}\right)  \label{output}%
\end{equation}

The solution is in register $X$. The oracle's choice has not been performed as
yet. It is performed by measuring $\left[  K\right]  $ in, indifferently,
(\ref{input}) or (\ref{output}). Say that we obtain $\ \mathbf{k}=01$. State
(\ref{output}) reduces on%

\begin{equation}
\frac{1}{\sqrt{2}}\left\vert 01\right\rangle _{K}\left\vert 01\right\rangle
_{X}\left(  \left\vert 0\right\rangle _{V}-\left\vert 1\right\rangle
_{V}\right)  . \label{am}%
\end{equation}

Measuring $\left[  X\right]  $\ in (\ref{am}) yields the solution produced by
the algorithm, namely the eigenvalue $\mathbf{x}$ $=01$.

In former work $\left[  5\right]  $, we showed that the quantum algorithm is
the sum over the (causal/local) histories of a classical algorithm that knows
in advance 50\% of the information about the solution. Each history
corresponds to a possible way of getting the advanced information (e. g., the
algorithm knows in advance that $k_{0}=0$) and to a possible result of
computing the missing information (e. g., the algorithm finds that $k_{1}=1$).
This decomposition of the quantum algorithm is the generalization of a well
known explanation of quantum nonlocality. We mean explaining the correlation
between the outcomes of two space-like separated quantum measurements by
connecting such outcomes with a causal/local history where causality is
allowed to go both forward and backward in time along the time reversible
quantum process. The following section provides the perfect classical machine
hidden in the quantum algorithm. The classical representation of quantum
retrocausation and nonlocality requires mechanical perfection: the hidden
machine should be perfectly rigid, accurate, and reversible. That infinite
classical precision can be dispensed for by quantization was already noted by
Finkelstein $\left[  11\right]  $.

\section{Many body interaction%
\index{cas1}
analogy}

The quantum data base search algorithm hides a perfect classical machine that
computes $\delta\left(  \mathbf{k},\mathbf{x}\right)  $ only once (the 2.25
computations on average apply to realistic, imperfect, classical machines).
This machine performs a hypothetical many body interaction that is actually a
visualization of the behavior of the qubit populations throughout quantum
measurement. This many body interaction representation shows that a
precondition of the quantum speed up is processing all the information
together at the same time.

We start with a representation of classical computation that highlights its
two body character. This is Fredkin\&Toffoli's billiard ball model of
reversible computation $\left[  12\right]  $. We have a billiard and a\ set of
balls moving and, from time to time, hitting each other or the sides of the
billiard, with no dissipation. We should prepare initial ball positions and
speeds so that there will be no many body collisions. This is not a problem,
it is just an essential feature of the machine: each individual collision is
between two balls or a ball and a side. Many body collisions should be avoided
because they yield undetermined outcomes -- this is the many body problem of course.

Where and when in this situation can we say that any amount of information is
processed together at the same time, as assumedly required to explain
perception? Outside collisions, the positions and speeds of different balls
are processed independently of one another. In collisions, the positions and
speeds of two balls are processed together at the same time. However, this
joint processing of information never scales up, it is always confined to ball
pairs. The information processed together at the same time is the three bits
of the input of a universal Boolean gate -- represented as a two body
interaction by Fredkin's controlled swap gate or Toffoli's
controlled-controlled not gate. Of course, parallel computation --\ several
two ball collisions at the same time -- does not count since the information
is not processed together. Summing up, we should discard classical computation
as a model of perception, because the amount of information processed together
at the same time is no more than three bits.

The many body problem arises when more than two balls collide together at the
same time. The problem is that the outcome of the collision is undetermined.
However, this is an idealization; in fact the slightest dispersion in the
times of pairwise collisions restores deterministic two body behavior.

Now we describe the perfect classical machine (perfectly rigid, accurate, and
reversible) hidden in the quantum database search algorithm -- see also
$\left[  2\right]  $ and $\left[  3\right]  $. We represent $\delta\left(
\mathbf{k},\mathbf{x}\right)  $, a function of the binary strings
$\mathbf{k}\equiv k_{0}k_{1}$ and $\mathbf{x}\equiv x_{0}x_{1}$, by the system
of Boolean equations%
\begin{align}
y_{0}  &  =\sim XOR\left(  k_{0},x_{0}\right)  ,\nonumber\\
y_{1}  &  =\sim XOR\left(  k_{1},x_{1}\right)  ,\nonumber\\
\delta\left(  \mathbf{k},\mathbf{x}\right)   &  =AND(y_{0},y_{1}),
\label{bequation}%
\end{align}

of truth tables%
\begin{equation}%
\begin{tabular}
[c]{|c|c|c|c|}\hline
& $k_{0}$ & $x_{0}$ & $y_{0}$\\\hline
$C_{00}$ & 0 & 0 & 1\\\hline
$C_{01}$ & 0 & 1 & 0\\\hline
$C_{02}$ & 1 & 0 & 0\\\hline
$C_{03}$ & 1 & 1 & 1\\\hline
\end{tabular}
\ \ ,%
\begin{tabular}
[c]{|c|c|c|c|}\hline
& $k_{1}$ & $x_{1}$ & $y_{1}$\\\hline
$C_{10}$ & 0 & 0 & 1\\\hline
$C_{11}$ & 0 & 1 & 0\\\hline
$C_{12}$ & 1 & 0 & 0\\\hline
$C_{13}$ & 1 & 1 & 1\\\hline
\end{tabular}
\ \ ,%
\begin{tabular}
[c]{|c|c|c|c|}\hline
& $y_{0}$ & $y_{1}$ & $\delta$\\\hline
$C_{20}$ & 0 & 0 & 0\\\hline
$C_{21}$ & 0 & 1 & 0\\\hline
$C_{22}$ & 1 & 0 & 0\\\hline
$C_{23}$ & 1 & 1 & 1\\\hline
\end{tabular}
\ \ . \label{truth}%
\end{equation}
The $C_{ij}$ ($i=0,~1,~2$, $j=0,~1,~2,~3$) labeling the rows of the truth
tables are real non-negative variables. They are the coordinates of the
machine parts -- our hidden variables. We replace the system of Boolean
equations (\ref{bequation}) by the following system of equations, representing
mechanical constraints between the coordinates of the machine parts,%

\begin{align}
\forall i  &  :Q=\sum_{j}C_{ij},\text{ \ \ \ }Q^{\chi}=\sum_{j}C_{ij}^{\chi
},\label{lin1}\\
C_{01}+C_{02}  &  =C_{20}+C_{21},\text{ \ \ \ }C_{11}+C_{12}=C_{20}+C_{22},
\label{lin3}%
\end{align}
with $\chi>1$. $Q$ is an auxiliary coordinate. In (\ref{lin1}), we can think
that left equations are implemented by three differential gears, one for each
truth table $i$. Each gear has one input $Q$ and four outputs $C_{i0}%
,C_{i1},C_{i2},C_{i3}$; right equations are implemented by a similar
arrangement with input $Q^{\chi}$ and outputs $C_{i0}^{\chi},C_{i1}^{\chi
},C_{i2}^{\chi},C_{i3}^{\chi}$, obtained from the former coordinates by means
of nonlinear transmissions. Equations (\ref{lin3}) are implemented by other
two differential gears, each with two inputs and two outputs, and the
coordinate $C_{20}$\ replicated in each gear.

We discuss the behavior of this analog computer assembling it step by step:

\begin{enumerate}
\item We start with one of the left equations/gears (\ref{lin1}), $Q=\sum
_{j}C_{ij}$, for some value of $i$. Initially all coordinates are zero. If we
push (the part of coordinate) $Q$, the $C_{ij}$ move to satisfy push and
equation. Collisions between bodies are replaced by pushing between
parts\footnote{Conversely, we could replace the billiard ball model of
classical computation by the present model, which can represent both many body
and two body interactions.}. A push instantly changes the force (or couple)
applied to a part from $0$ to $\neq0$. The outcome of this many body
interaction is undetermined: for a given $Q$,\ there are infinitely many
possible "machine movements". We have a many body interaction between $4$
machine parts of coordinates $C_{ij}$ -- choosing $Q$ as the dependent
variable. Since we have to match machine behavior with the transition from
state (\ref{output}) before measurement\ to one of four possible states after
measurement, each occurring with probability $\frac{1}{4}$, we postulate that
the probability distribution of the machine movements is symmetrical for the
exchange of any two $C_{ij}$.

\item We add the right equation/gear, $Q^{\chi}=\sum_{j}C_{ij}^{\chi}$, for
the same value of $i$. Now pushing $Q$ can\ move at most one $C_{ij}$ --
$C_{ij}$ movements become mutually exclusive with one another. Perfect
coincidence of the times of the push exchanged between parts requires perfect
rigidity and accuracy of the machine. Flexibility and other imperfections
restore deterministic two body behavior, likely with an ordering of pairwise
pushes that frustrates the mechanical constraints, thus jamming the machine.
For example, if two or more $C_{ij}$ move initially, thanks to a slight
deformation of the mechanical constraints, the further movement of $Q$
increases the deformation until the machine jams. No deformation, i. e.
machine perfection, implies no jams, namely postulating that one of the
$C_{ij}$ moves to satisfy push and equations. Symmetry of the probability
distribution yields even probabilities of movement for the $C_{ij}$.\ The
machine movement produces the Boolean values of the row (of the truth table
$i$)\ labeled by the one $C_{ij}>0$.

\item We add the remaining equations/gears. Equations (\ref{lin1}) assure that
only one $C_{ij}$\ moves for each $i$, equations (\ref{lin3}) assure that the
$C_{ij}$ that move label the same values of the same Boolean variables, namely
that the machine movement satisfies the system of Boolean equations
(\ref{bequation}).

\item If we push $Q$, there are $16$ mutually exclusive machine movements,
corresponding to all the possible ways of satisfying the system of Boolean
equations (\ref{bequation}). We have a many body interaction between the $8$
machine parts of coordinates $C_{0j}$ and $C_{1j}$, the other coordinates
being dependent variables.

\item If we push $C_{23}$\ instead of $Q$, the movement of $C_{23}$ yields
$\delta\left(  \mathbf{k},\mathbf{x}\right)  =1$. Now there are $4$ mutually
exclusive machine movements. Each movement produces an oracle's choice and the
corresponding solution provided by the second player by means of a single
computation of $\delta\left(  \mathbf{k},\mathbf{x}\right)  $ -- a single
transition $C_{23}=0\rightarrow C_{23}>0$.
\end{enumerate}

This latter many body interaction represents the behavior of the qubit
populations throughout quantum measurement. In fact there is an invertible
linear relation between the eight $\frac{C_{0j}}{Q}$,$\frac{C_{1j}}{Q}$
\ ($j=0,1,2,3$) and the eight\ qubit populations. For example, with reference
to the reduced density operator of qubit $k_{0}$, let $p_{k_{0}}^{00}$ be the
population of $\left\vert 0\right\rangle _{k_{0}}\left\langle 0\right\vert
_{k_{0}}$, and $p_{k_{0}}^{11}$ that of $\left\vert 1\right\rangle _{k_{0}%
}\left\langle 1\right\vert _{k_{0}}$. By looking at the truth tables, one can
see that their relation with the $\frac{C_{ij}}{Q}$ is:%

\begin{equation}
p_{k_{0}}^{00}=\frac{C_{00}+C_{01}}{Q},~p_{k_{0}}^{11}=\frac{C_{02}+C_{03}}%
{Q}. \label{populations}%
\end{equation}
The relation for the other qubits, $k_{1}$, $x_{0}$, and $x_{1}$, is derived
in a similar way. When all coordinates are $0$, all ratios are $\frac{0}{0}%
$\ and are thus compatible with any value of the populations in the state
before measurement. Having postulated a symmetric probability distribution of
machine movements sets to $\frac{1}{2}$ the values of the qubit populations
before measurement (like in state \ref{output}). When $C_{23}>0$, these ratios
become determined and correspond to either 0's or 1's of the populations of
the measured observables: the $C_{ij}$\ that do not move yield $\frac{C_{ij}%
}{Q}=0$, those that move yield $\frac{C_{ij}}{Q}=1$.

This many body analogy helps to understand what goes on, computationally, in
quantum measurement: satisfaction "in one shot" -- with a single computation
of $\delta\left(  \mathbf{k},\mathbf{x}\right)  $ -- of the nonlinear system
of Boolean\ equations constituted by (\ref{bequation}) and $\delta\left(
\mathbf{k},\mathbf{x}\right)  =1$ (satisfied by pushing $C_{23}$).

On the contrary, satisfying this system classically, by means of deterministic
two body interactions, would require on average, 2.25 computations of
$\delta\left(  \mathbf{k},\mathbf{x}\right)  $. More in general, a classical
computation satisfies in one shot (i. e. satisfying each gate at the first
attempt) a linear Boolean network, in fact through the deterministic
propagation of an input into the output. In the case of a nonlinear network,
local deterministic satisfaction of gates can be done in several ways, and is
likely done in a way that does not satisfy other gates. This leads to trial
and error and repeated computations, which yields the relative zero of the
quantum speed up.

In the initial state of the quantum algorithm (\ref{input}), the hidden
machine is disassembled and the coordinates of the machine parts are
independent of one another. Correspondingly the quantum state is factorizable
-- quantum measurement of the register contents would yield uncorrelated outcomes.

The unitary part of the quantum algorithm, yielding state (\ref{output}),
assembles the machine: all parts -- in the configuration all coordinates zero
-- get geared together in a non-functional relation (established by equations
\ref{lin1}, \ref{lin3}). Correspondingly the quantum state is entangled.
Measuring the register contents in this state corresponds to operating the
machine -- to pushing $C_{23}$. This generates the interaction that in one
shot produces the oracle's choice, runs the algorithm, and produces the solution.

This many body analogy can easily be generalized.

If several function evaluations are required, like in data base search with
$N>4$, just one computation of $\delta\left(  \mathbf{k},\mathbf{x}\right)  $
and one rotation of the $X$\ basis creates the superposition of a state of
maximal entanglement between $K$ and $X$\ (corresponding to the assembled
machine) and the factorizable initial state back again $\left[  4\right]
,\left[  5\right]  $ (corresponding to the disassembled machine). Iterating
these operations $\operatorname{O}\left(  \sqrt{N}\right)  $ times "pumps" the
amplitude of the entangled state to about 1. Measurement should be performed
-- the machine operated -- in this final state.

In the other quantum algorithms, the oracle chooses a function $f_{\mathbf{k}%
}\left(  x\right)  $ out of a known set of functions and gives to the second
player the black box for its computation. The second player should find out a
certain property of the function (e. g. its period) by means of one
computation of $f_{\mathbf{k}}\left(  x\right)  $ -- against, classically, a
number of computations exponential in the size of the argument. It is
sufficient to: (i) represent the oracle's choice and the property of the
function as a network of Boolean gates, with the rows of the truth tables
labeled by the hidden variables, (ii) introduce the equivalent system of
equations on the qubit populations (iii) assemble the perfect machine through
the unitary evolution part of the quantum algorithm, and (iv)\ operate it by
measuring the register contents. Quantum measurement satisfies in one shot a
nonlinear Boolean network.

\section{Interdisciplinary implications}

The notion that a quantum algorithm knows in advance 50\% of the solution it
will find in the future, and the related notion of satisfying in one shot a
nonlinear Boolean network, interplay with a variety of scientific and
philosophical issues. In the following, we call the many body interaction
hidden in the measurement stage of the quantum algorithms \textit{simultaneous
computation}.

Among the scientific issues, we find:

\begin{itemize}
\item The character of visual perception implies the capability of processing
together at the same time a significant amount of information. Simultaneous
computation can process in this way any amount of information, therefore it
can be the physical computation basis of perception. Classical computation,
capable of processing together at the same time no more than three bits, could not.

\item Simultaneous computation provides a formalization of the physical
computation level of those neurophysiological and physical theories that place
consciousness in quantum measurement, like Hameroff\&Penrose's orchestrated
objective reduction theory $\left[  16\right]  ,\left[  17\right]  ,\left[
18\right]  $ and Stapp's theory $\left[  25\right]  $.

\item Let us adopt the strong artificial intelligence (AI) assumption that a
state of consciousness is a computation process with an upper bound to the
number of computation steps, thus representable as the process of satisfying a
Boolean network. In the present perspective, the entire computation should be
performed in one shot, together at the same time, by quantum measurement. To
match subjective experience, the computation should represent the feeling of
self, memories, emotion, thinking, sensorial information, etc. Most of the
processing (e. g. the feeling of self) would be repeated at each successive
measurement; part of the processing would be updated to track changes -- in
memories, emotions, etc. A frequency of 50 measurements per second (50 "frames
per second"), could cope with our rates of change.

\item Simultaneous computation solves -- at the physical computation level --
the "hard problem" pinpointed by Chalmers $\left[  7\right]  $: explaining how
disparate informations can come together in the unity of subjective experience
-- this unity is processing together at the same time all information.

\item A qualia is an atomic sensation -- apparently without an internal
logical structure -- like that of "redness" -- see Ref. $\left[  22\right]  $.
Classical computation is phenomenological in character, feeling a qualia would
correspond to an algorithm that behaves consistently with that feeling
(talking of the red color, stopping at a red light). In the context of quantum
simultaneous computation, "seeing", or "feeling", are synonyms of "measuring".
Feeling a qualia could correspond to measuring some fundamental observable
(and, at the same time, the self -- possibly comprising other qualia -- and
some relation between feeling of self and the feeling of a color).

\item Identifying consciousness with simultaneous computation -- i. e. the
mechanism enabling the quantum speed up -- gives quantum consciousness a
potential evolutionary advantage over a classical consciousness. The former
could be immensely quicker and/or leaner in computational resources in tasks
essential for survival. With respect to classical computation, quantum
associative memory requires an exponentially lower number of artificial
neurons $\left[  28\right]  $, quantum pattern recognition can be traced back
to quantum data base search, which yields a quadratic speed up $\left[
27\right]  ,\left[  30\right]  $, quantum machine learning has recently been
shown to be substantially faster $\left[  20\right]  $.

\item Teleological evolutions often explain organic behavior better than
deterministic classical evolutions -- see Ref. $\left[  13\right]  $. However,
such explanations are generally considered to be phenomenological in
character, because of the belief that, really, evolutions could not be driven
by final conditions. Quantum algorithms, being partly driven by their future
outcome, provide well formalized examples of teleological evolutions.

\item Stapp's theory relies on the quantum Zeno effect and lives with
decoherence -- see Ref. $\left[  25\right]  $. The present model suffers
decoherence exactly as quantum computation does, which means very much. This
divergence could mean cross fertilization. It puts emphasis on the quantum
information approach of driving the state of the computer registers by means
of the Zeno effect -- see Ref. $\left[  22\right]  $.

\item The notion that quantum algorithms are partly driven by their future
outcome is consistent with Sheehan's retrocausation theory and critical
revision of the notions of time and causality in physics -- see Ref. $\left[
23\right]  $.
\end{itemize}

Among the philosophical issues, we find:

\begin{itemize}
\item Being entirely driven by past conditions excludes free will, as well as
being entirely driven by future conditions. Being partly driven by either
condition -- like quantum algorithms -- leaves room for freedom. In quantum
algorithms, freedom from determinism is nondeterministic computation --
capability of satisfying in one shot a nonlinear Boolean network.

\item A quantum algorithm, for the fact of knowing in advance 50\% of the
solution it will find in the future, "exists" in an extended present. This
suggests that our existence is not confined to the instantaneous present we
normally experience. With reference to Indian philosophy, the experience of an
instantaneous present would be illusory, the timeless reality experienced in
Moksa (in western language, in special "altered states of consciousness" --
see Ref. $\left[  8\right]  $)\ objective.

\item Insight -- understanding an even immensely complex structure in one
instant -- seems to be a most evident experience of simultaneous computation.

\item Simultaneous computation has an upper bound to the number of computation
steps, like quantum algorithms and AI. This is a limitation with respect to
Lucas-Penrose's argument that consciousness -- being able to "see" G\"{o}del's
theorems -- is not confined to finitistic computation -- see Ref. $\left[
19\right]  ,\left[  21\right]  $. As for the possibility of extending
simultaneous computation to denumerably infinite Boolean networks, see Ref.
$\left[  6\right]  $.

\item Mind-body duality, or the duality between a perfect world of ideas and
an imperfect material world, here becomes the duality between (i)
perfect/nondeterministic classical machines (hidden in quantum measurement),
yielding a speed up and capable of processing any amount of information
together at the same time, thus of hosting consciousness, and
(ii)\ imperfect/deterministic classical machines, capable of processing no
more than three bits together at the same time, incapable of hosting
consciousness. This also matches with Stapp's distinction between the mind and
the rock aspect of quanta $\left[  25\right]  $. If there is only quantum
physics, this duality vanishes. The perfect/nondeterministic side would be
objective, the other side phenomenological or illusory.
\end{itemize}

\section{Conclusions}

The advanced knowledge of the solution, which explains the quantum speed up,
has been seen as a many body interaction between the parts of a perfect
classical machine whose coordinates represent the qubit populations throughout
quantum measurement. In one shot (with a single input-output transformation of
each gate), this interaction senses and satisfies all the gates of a nonlinear
Boolean network together at the same time.

In contrast, the amount of information processed together at the same time by
classical computation is limited to the three bit input of a single universal
Boolean gate -- many such gates in parallel do not count since the information
is not processed together. Correspondingly, classical computation cannot
satisfy a nonlinear Boolean network in one shot (but for a very lucky instance).

Simultaneous computation answers our prerequisite for the physical computation
level of perception -- capability of processing any amount of information
together at the same time. With reference to the theories that place
consciousness in quantum measurement, simultaneous computation takes two
pigeons with one stone:

\begin{enumerate}
\item it formalizes the physical computation level of these theories,

\item in such a way that the fundamental mechanism of consciousness is the
same of the quantum speed up.
\end{enumerate}

The overall result is giving quantum consciousness, with respect to classical
consciousness, a potentially enormous evolutionary advantage.

More in general, simultaneous computation could be the physical computation
level of biological information processing. It\ provides a scientific ground
to teleological explanations of organic behavior and a possible answer to long
standing philosophical questions.

The assumption that biological computation is simultaneous computation implies
that the brain hosts a sufficient quantum coherence -- see Ref. $\left[
10\right]  ,\left[  15\right]  ,\left[  25\right]  ,\left[  26\right]
,\left[  29\right]  $. It can be argued that the problem of decoherence is
common to quantum information, whose alleged advantage -- possibility of
working close to 0 Kelvin and without hydrophobic pressure -- is frustrated by
the fact that the size of the computation cannot scale up in any conceivable
way. Our biased opinion is that the top level evidence that the mind is
quantum, and cannot be classical, is strong enough to look for a common
solution. Tackling the problem of decoherence from the two leads -- quantum
information and biological -- might yield cross fertilization.

\subsection*{Acknowledgements}

I thank for useful discussions, Vint Cerf, Artur Ekert, David Finkelstein,
Shlomit Finkelstein, Hartmut Neven, Barry Wessler, and my wife Ferdinanda.

\subsection{Bibliography}

$1.$ Castagnoli, G.: The mechanism of quantum computation. Int. J. Theor.
Phys.,\textit{\ }vol. \textbf{47}, number 8, 2181 (2008)

$2.$\ Castagnoli, G.: The quantum speed up as advanced cognition of the
solution. Int. J. Theor. Phys., vol. \textbf{48}, issue 3, 857 (2009)

$3.$\ Castagnoli, G.: The 50\% advanced information rule of the quantum
algorithms. Int. J. Theor. Phys. vol. 48, issue 8, 2412 (2009)

$4.$ Castagnoli, G.: Quantum algorithms know in advance 50\% of the solution
they will find in the future. Int. J. Theor. Phys.,vol. 48 issue 12, 3383
(2009) DOI:10.1007/s10773-009-0143-6, arXiv:quant-ph/009008? v4

$5$. Castagnoli, G.: Explanation of the quantum speed up, submitted to Pys.
Rev. A (2009)

$6.$ Castagnoli, G., Rasetti, M., Vincenzi, A.: Steady, simultaneous quantum
computation: a paradigm for the investigation of nondeterministic and
non-recursive computation. Int. J. of Mod. Phys. C, 3, No. 4, 661 (1992)

$7.$ Chalmers, D.: Facing Up the Problem of Consciousness. Journal of
Consciousness Studies, 2, 200-219 (1995)

$8.$ De Faccio, A.: From an altered state of consciousness to a life long
quest of a model of mind. TASTE Archives of Scientists' Transcendent
Experiences, submission N\ 00098. Charles T. Tart editor.

http://www.issc-taste.org/arc/dbo.cgi?set=expom\&id=00088\&ss=1 (2002)

$9.$ Deutsch, D.: Quantum theory, the Church-Turing principle, and the
universal quantum computer. Proc. Roy. Soc. (Lond.) A, \textbf{400}, 97 (1985)

$10.$ Engel, G. S., Calhoun, T. R., Read, E. L., Ahn, T. K., Mencal, T.,
Cheng, Y. C., Blankenship, R. E., and Fleming, G. R.: Evidence for wavelike
energy transfer through quantum coherence in photosynthetic systems. Nature,
446, 782 (2007)

$11.$\ Finkelstein, D. R.: Generational Quantum Theory. Preprint, to become a
Springer book (2008)

$12.$ Fredkin, E. and Toffoli, T. Conservative logic.\textit{\ Int. J. Theor.
Phys.} 21, 219 (1982)

$13.$ George, F. H. and Johnson, L.: Purposive Behaviour and Teleological
Explanations. Studies in Cybernetic, vol. 8. Gordon And Breach Science
Publishers (1985)

$14.$ Grover, L. K.: A fast quantum mechanical algorithm for data base search.
Proc. 28th Ann. ACM Symp. Theory of Computing (1996)

$15.$ Hagan, S., Hameroff, S. R., and Tuszynski,J. A.: Quantum Computation in
Brain Microtubules? Decoherence and Biological Feasibility. Physical Reviews
E, vol. 65, 061901 (2002)

$16.$ Hameroff, S. R.: The Brain Is Both Neurocomputer and Quantum Computer.
Cognitive Science 31, 1035-1045 (2007)

$17.$ Hameroff, S. R.: The "conscious pilot" - dendritic synchrony moves
through the brain to mediate consciousness. Journal of Biological Physics. http://www.springerlink.com/content/?k=10.1007/s10867-009-9148-x

$18.$ Hameroff, S. R. and Penrose, R.: Toward a Science of Consciousness. The
First Tucson Discussions and Debates, eds. Hameroff, S. R., Kaszniak, A. W.,
and Scott, A. C., Cambridge, MA: MIT Press, 507-540 (1996)

$19.$ Lucas J. R.: The Godelian Argument.
http://www.leaderu.com/truth/2truth08.html (July, 2002)

$20.$ Neven, H., Dencher, V. S., Rose, G., and Macready, W. G.: Training a
Binary Classifier with the quantum Adiabatic Algorithm. arXiv 0811.0416v1

[quant-ph] (2008)

$21.$ Penrose, R.: Shadows of the Mind -- a Search for the Missing Science of
Consciousness. Oxford University Press (1994)

$22.$ Searle, J. R.: Mind, a Brief Introduction. Oxford University Press (2004)

$23.$ Shehan, D. P.: Frontiers of Time: Retrocausation -- Experiment and
Theory, San Diego, California, 20-22 June 2006

$24.$ Sh\"{u}lte-Herbr\"{u}ggen, T., Sp\"{o}rl, A., Khaneja, N., Glaser, S.
J.: Optimal Control for Generating Quantum Gates in Open Dissipative Systems.
arxiv:quant-ph/0609037 (2009)

$25.$ Stapp, H. P.: Mind Matter and Quantum Mechanics. Springer (March 2009)

$26.$ Summhammer, J., Bernroider, G.: Quantum entanglement in the voltage
dependent sodium channel can reproduce the salient features of neuronal action
potential initiation. arXiv:0712.1474v1[physics.bio-ph] (2007)

$27.$ Trugenberger, C. A.: Quantum Pattern Recognition.
arXiv:quant-ph/0210176v2 (2002)

$28.$ Ventura, D. and Martinez, T.: Quantum Associative Memory. Information
Sciences, vol. 124, nos 1-4, 273-296 (2000)

$29.$ Vitiello G.: Coherent States, Fractals, and Brain Waves. New Mathematics
and Natural Computing, vol. 5, N. 1, 245-264 (2009)

$30.$ Zhou, R. and Ding, Q.: Quantum Pattern Recognition with Probability
100\%. Int J. Theor. Phys., vol.47, N. 5 (2008)

\end{document}